\documentclass[12pt]{iopart}

\usepackage[english]{babel}
\usepackage{enumerate}
\usepackage{multicol}
\usepackage{fancyhdr}
\usepackage{amssymb}
\usepackage{hhline}
\usepackage{wrapfig}
\usepackage{caption}
\usepackage[dvips]{graphicx}
\usepackage{pifont}
\usepackage{makeidx}
\usepackage{cite}
\usepackage[T1]{fontenc}

\begin{document}
\title[New features in the phase diagram of TbMnO$_3$]{New features in the phase diagram of TbMnO$_3$}
\author{D~Meier$^{1,2}$, N~Aliouane$^{3}$, D~N~Argyriou$^{3}$, J~A~Mydosh$^{1}$ and T~Lorenz$^{1}$}
\address{$^1$ II. Physikalisches Institut, University of Cologne, Z\"{u}lpicherstrasse 77, D-50937 K\"{o}ln, Germany}
\address{$^2$ Helmholtz-Institut f\"{u}r Strahlen- und Kernphysik, University of Bonn,
Nu\ss allee 14-16, D-53115 Bonn, Germany}
\address{$^3$ Hahn-Meitner-Institut Berlin, Glienicker Str. 100, D-14109 Berlin, Germany}
\ead{meier@ph2.uni-koeln.de}
\begin{abstract}
The $(H,T)$-phase diagram of the multiferroic perovskite
TbMnO$_3$ was studied by high-resolution thermal expansion
$\alpha(T)$ and magnetostriction $\Delta L(H)/L$ measurements.
Below $T_{N}\simeq 42$~K, TbMnO$_3$ shows antiferromagnetic
order, which changes at $T_{\rm FE}\simeq 28$~K where
simultaneously a spontaneous polarization $P||c$ develops.
Sufficiently high magnetic fields applied along $a$ or $b$ induce a
polarization flop to $P||a$. We find that all of these
transitions are strongly coupled to the lattice parameters. Thus,
our data allow for a precise determination of the phase
boundaries and also yield information about their uniaxial
pressure dependencies. The strongly hysteretic phase boundary to
the ferroelectric phase with $P||a$ is derived in detail.
Contrary to previous reports, we find that even in high magnetic
fields there are no direct transitions from this phase to the
paraelectric phase. We also determine the various phase
boundaries in the low-temperature region related to complex
reordering transitions of the Tb moments.
\end{abstract}


 \maketitle

\section{Introduction}

The coupling between magnetic and ferroelectric order parameters
in so-called magnetoelectric multiferroic materials is of great
current
interest~\cite{eerenstein06a,tokura06a,khomskii06a,heyer06a,senff06a}.
After the rediscovery of this mechanism a few years ago, an
intense search has started for materials with strong coupling of
the spontaneous polarization $P$ and the spontaneous
magnetization $M$~\cite{spaldin05a,fiebig05a,hill}. This search
is enhanced, on the one hand, by the demand for new promising
components for device design~\cite{hur04a}. On the other hand, the
fundamental aspects of the magnetoelectric coupling in many
systems are far from being understood. Thus, a detailed
determination of the complex and rich phase diagrams observed in
multiferroic materials is important. Considering the multiferroic
orthorhombic manganites \textit RMnO$_3$ (\textit R = Gd, Tb,
Dy), most of the previous investigations concentrated on the
ordering phenomena related to the Mn ions, while less is known
about the ordering of the rare earth ions in magnetic
fields~\cite{kimura03b,kajimoto04a,aliouane05a}. Recently, we
have shown that measurements of thermal expansion and
magnetostriction by high-resolution dilatometry are a powerful
method to investigate the temperature and magnetic-field phase
diagrams of multiferroics, because both the magnetic and the
ferroelectric order strongly couple to the lattice
parameters~\cite{baier06a}. These couplings reflect pronounced
uniaxial pressure dependencies of the respective transition
fields and temperatures, which can be analyzed by
Clausius-Clapeyron and Ehrenfest relations for first- and
second-order phase transitions, respectively~\cite{baier06b}.

Below the N\'{e}el temperature $T_{\rm N}\simeq 42$~K the Mn spins of
TbMnO$_3$ develop an incommensurate sinusoidal antiferromagnetic
alignment with wave vector $(0,k_{Mn}(T),0)$~\cite{kimura05a}
(using the Pbnm setting of the orthorhombic unit cell). According
to Harris {\it et al.}~\cite{harris06a}, this phase is called the
high-temperature incommensurate antiferromagnetic phase (HTI). At
$T_{\rm FE}\simeq 28$~K another transition occurs, leading also
to an incommensurate, but cycloidal ordering of the Mn moments
along $(0,k_{Mn}(T),0)$. Because the latter transition breaks the
inversion symmetry of the crystal, a spontaneous polarization
$P||c$ can appear below $T_{\rm FE}$ even in zero field
\cite{mostovoy06a,katsura05a}. This phase is called the
low-temperature incommensurate AFM phase (LTI). In zero field, an
incommensurate AFM ordering of the Tb ions has been
proposed~\cite{kajimoto04a}, which sets in below $T_{\rm N}^{\rm
Tb}=7$~K. The above-described ordering phenomena of the Mn ions
hardly change for magnetic fields $H||i <  4.5$~T ($i=a$, $b$,
$c$). However, in higher magnetic fields significant changes in
the ferroelectric LTI phase occur: sufficiently high fields
($H>H_{\rm FE,C}$) parallel to the $a$ or $b$ axis induce a
polarization flop from $P||c$ to $P||a$. The critical field
strengths $H_{\rm FE,C}$ depend on their orientation and on
temperature. In general, $H_{\rm FE,C}$ is smaller for $H||b$
($H^{b}_{\rm FE,C}\gtrsim 4.5$~T) than for $H||a$ ($H^{a}_{\rm
FE,C}\gtrsim 9.5$~T)~\cite{aliouane05a}. The polarization flop is
accompanied by a change in the modulation wave vector, which
becomes commensurate for $H > H_{\rm FE,C}$ (LTC
phase)~\cite{kimura05a}. If a magnetic field $H \gtrsim 8$~T is
applied parallel to the $c$ axis, the sample is forced into a
canted antiferromagnetic ordering (cAFM phase) and the
ferroelectric order is suppressed completely~\cite{aliouane05a}.

In this paper we present a study of the linear thermal expansion
coefficient \mbox{$\alpha_i(T)=\frac{1}{L_i} \frac{\partial
L_i(T)}{\partial T}$} and the linear magnetostriction
$\frac{\Delta L_i(H)}{L_i}=[L_i(H)-L_i(0)]/L_i(0)$ of TbMnO$_3$.
Here, $L_i$ denotes the length of the sample parallel to the
different crystallographic directions $i =a$, $b$, and $c$. The
TbMnO$_3$ single crystal was cut from a larger crystal grown by
floating-zone melting in an image furnace~\cite{aliouane05a}. The
paper is organized as follows: in the next section we present
zero field measurements of the linear thermal expansion
coefficients. Section~3 deals with the phase transitions which
are attributed to the Mn sublattice. We performed measurements
along all three lattice directions in longitudinal magnetic
fields up to $H=16$~T, i.e., in all cases the field was applied
parallel to the measured crystal axis ($H||i$ with $i=a$, $b$,
and $c$). The phase diagram derived from our data and the
uniaxial pressure dependencies of various phase boundaries are
discussed in the fourth section. Section~5 considers the ordering
transitions related to the Tb moments and the last section gives
a summary.

\section{Linear thermal expansion in zero field}

\begin{figure}
    \centering
        \includegraphics[width=0.80\textwidth]{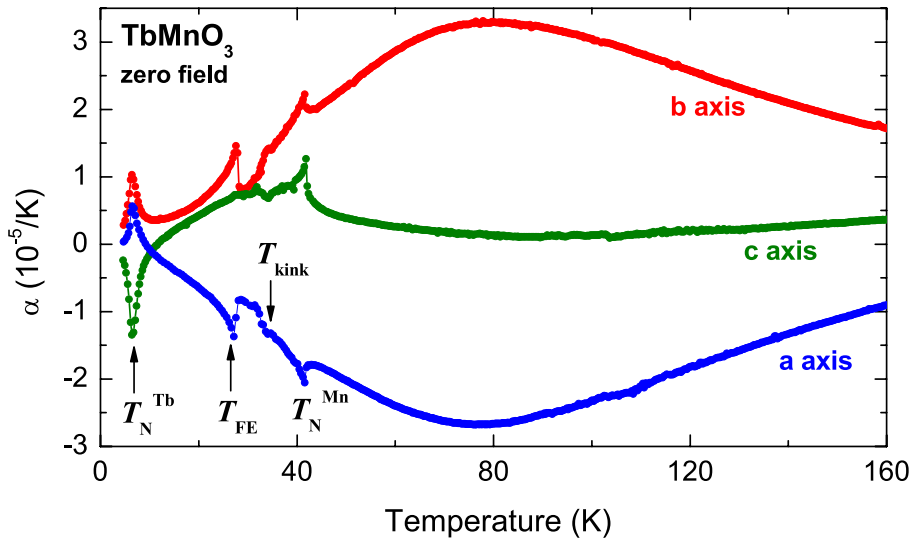}
        \caption{Thermal expansion of TbMnO$_3$ measured along
        the $a$, $b$, and $c$ axis.}
    \label{fig:zerofieldcut}
\end{figure}

The thermal expansion coefficients $\alpha_i(T)$ of TbMnO$_3$
measured in zero magnetic field are shown in
figure~\ref{fig:zerofieldcut}. As may be expected from the
orthorhombic structure, the TbMnO$_3$ single crystal shows a
strongly anisotropic thermal expansion. Several anomalies are
detected. These anomalies can be attributed to the different phase
transitions according to previous publications~\cite{kimura05a}.
The sharp anomalies at $41.5$~K signal the N\'{e}el transition of the
Mn ions. At $T_{\rm FE}=27.6$~K TbMnO$_3$ undergoes a second phase
transition from the HTI to the LTI phase, accompanied by a
pronounced anomaly in $\alpha_i(T)$ for $i=a$ and $b$, whereas
the effect in $\alpha_c(T)$ is much smaller. The shape of the
anomalies at $T_{\rm N}$ and $T_{\rm FE}$ is typical for
second-order phase transitions. The uniaxial pressure dependencies
of $T_C$ are described by the Ehrenfest relation
\begin{eqnarray}
\label{Ehr}
 \frac{\partial T_C}{\partial p_i} = V_m
T_C\frac{\Delta \alpha_i}{\Delta c}  \ .
\end{eqnarray}
Here, $V_m$ is the molar volume, $\Delta \alpha_i$ ($\Delta c$)
denotes a jump in the thermal expansion coefficient (specific
heat) and $i$ is the measurement direction of $\alpha_i$ and the
direction of uniaxial pressure $p_i$. Because any ordering
transition causes a decrease of the entropy, the specific heat
anomaly is always positive ($\Delta c
> 0$) and the sign of $\partial T_C /\partial p_i$ is given by
the sign of $\Delta \alpha_i$. Thus, $T_{\rm N}$ shifts to higher
temperature for uniaxial pressure $p_i$ along the $b$ or $c$ axis
and decreases for $p_a$. The changes of $T_{\rm FE}$ for $p_a$
and $p_b$ have the same signs as those of $T_{\rm N}$, i.e.,
$T_{\rm FE}$ decreases for $p_a$ and increases for $p_b$.
Pressure parallel to the $c$ axis has only a minor effect on
$T_{\rm FE}$, since practically no change $\Delta \alpha_c$ is
observed at $T_{\rm FE}$.
The ordering of the Tb sublattice at $T_{\rm N}^{\rm Tb}=7$~K also
causes pronounced anomalies for all three crystallographic
directions. From the signs of the thermal expansion anomalies we
conclude that the Tb ordering is stabilized for uniaxial pressure
along $a$ or $b$, while it is suppressed for $p_c$.

Apart from the anomalies related to the different transition
temperatures there are additional anomalous features visible for
all three directions. Firstly, kink-like anomalies appear at
$T_{\rm kink}\simeq 34$~K. We suspect that these kinks may be
related to the slope change in the temperature dependence of
$k_{Mn}(T)$ reported by Kajimoto {\it et al.}, although the latter
has been observed at a somewhat larger temperature $T\simeq
38$~K~\cite{kajimoto04a}. Secondly, $\alpha_a$ ($\alpha_b$) shows
a pronounced broad minimum (maximum) centered around $80$~K and a
weaker minimum is also present in $\alpha_c$. Such broad extrema
are typical indications for Schottky contributions arising from a
thermal population of low-lying excited states. The fact that
this thermal population contributes to the thermal expansion
coefficients reflects strong uniaxial pressure dependencies of
the relevant energy splitting(s) between the ground state and the
excited state(s). Again, the signs of the uniaxial pressure
dependencies are given by the signs of the anomalous
$\alpha_i$~\cite{zobel02a,johannsen06a,lorenz06a}. In TbMnO$_3$
these Schottky contributions most probably arise from the $4f$
multiplet of the Tb$^{3+}$ ions. The Hund's rule ground state of
the free Tb$^{3+}$ ion has a total orbital momentum $J=6$ and is
13-fold degenerate. In a crystal field of orthorhombic symmetry
this degeneracy is completely lifted and it is obvious that the
energy splittings between the 13 singlet states may strongly
change with pressure, since the crystal field will depend on
pressure. A more quantitative analysis of the Schottky
contributions to $\alpha_i$ is not possible at the present stage,
since it would require a detailed knowledge of the different
singlet states and of their energy splitting.

\section{Transitions related to the Mn ordering}

We will divide the discussion of our data in two parts. In this
section we will concentrate on the phase transitions related to
the Mn sublattice. The ordering of the Tb ions is discussed in
section~\ref{Tbordering}. This division does not mean that the Mn
and the Tb sublattice of TbMnO$_3$ act independently from each
other, but it is reasonable to assume that the coupling between
the two sublattices is not too strong.

\subsection{Measurements in magnetic fields $H||a$}

\begin{figure}
    \begin{center}
        \includegraphics[width=0.90\textwidth]{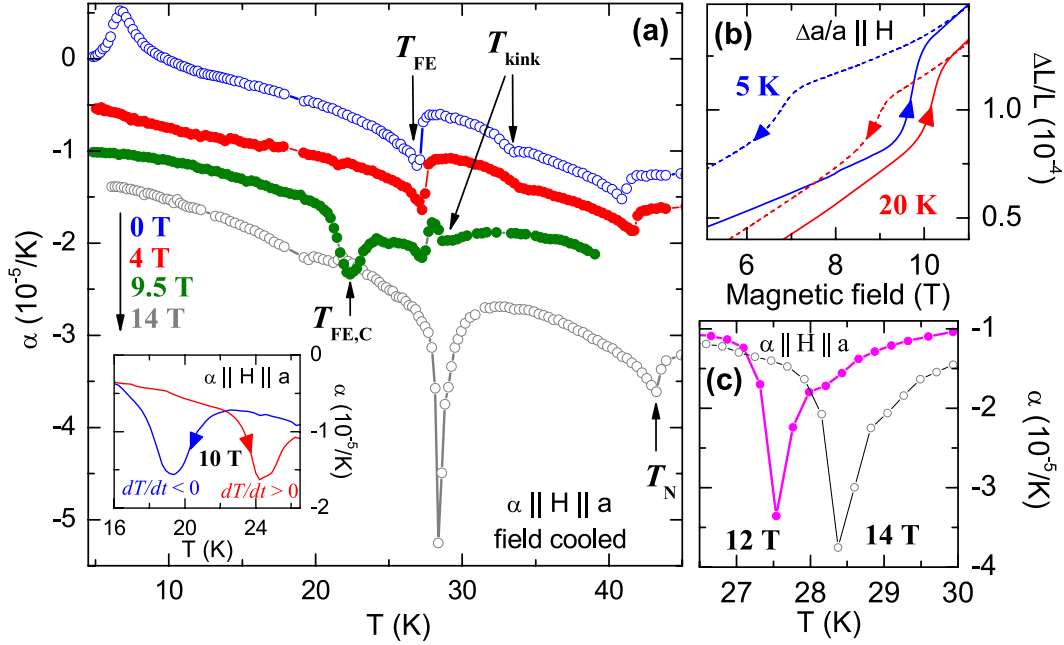}
    \end{center}
    \caption{Panel (a): Field-cooled (FC)
    thermal expansion $\alpha_a(T)$ for $H||a$ measured with increasing
    temperature. For clarity the curves for different fields are
    shifted by $-5\cdot 10^{-6}$/K with respect to each other. We find a
    pronounced hysteresis around $T_{\rm FE,C}$. This is exemplified
    in the inset by the $\alpha_a(T)$ curves measured in a field of 10~T
    with increasing and decreasing temperature. This hysteresis is
    also seen in magnetostriction measurements
    $\Delta a(H)/a$ recorded with increasing (solid lines) and
    decreasing (dashed lines) field shown in panel~(b). Panel~(c) displays an
    enlarged view on the high-field $\alpha_a(T)$ anomalies, which both
    have additional shoulders on their high-temperature sides, indicating the
    presence of two transition temperatures $T_{\rm FE,C}$ and $T_{\rm FE}$.}
    \label{fig:alphaa}
\end{figure}
Measurements of $\alpha_a(T)$ in a longitudinal magnetic field
$H||a$ up to 14~T are shown in figure~\ref{fig:alphaa}~(a). Above
about 10~K the $\alpha_a(T)$ curves in 4~T and in zero field are
quite similar. For higher magnetic fields $H\geq H^{a}_{\rm
FE,C}\simeq 9.5$~T an additional anomaly appears at $T_{\rm FE,C}$
which signals the phase transition from the LTI ($P||c$) to the
LTC phase with $P||a$. This first-order transition shows a broad
hysteresis. The hysteretic behavior is presented in the inset of
figure~\ref{fig:alphaa}~(a) by measurements of $\alpha_a(T)$ in
10~T with increasing ($dT/dt > 0$) and decreasing ($dT/dt < 0$)
temperature. The LTI-to-LTC transition can also be detected by
measurements of the magnetostriction, i.e.\ the magnetic-field
induced length change at a constant temperature. In
figure~\ref{fig:alphaa}~(b) magnetostriction measurements at
$T=5$~K and 20~K are presented. In both curves a jump-like
expansion of $\Delta a(H)/a$ is observed as the phase boundary
between the LTI and LTC phase is crossed as a function of
increasing $H$. For decreasing $H$, the anomalous expansion is
reversed and a jump-like contraction signals the LTC-to-LTI
transition. As in $\alpha_a(T)$, this first-order transition
shows a hysteresis, which is strongly enhanced at lower
temperature.

As is also seen in figure~\ref{fig:alphaa}~(a), $\alpha_a(T)$ in
$H=14$~T shows a pronounced anomaly at $\simeq 28.5$~K. At first
glance, the peak-like shape and a small hysteresis (not shown)
indicate that there is one first-order phase transition at this
temperature. However, a closer inspection of the high-field
anomalies for $H=12$~T and 14~T reveals that both anomalies are
asymmetric with additional shoulders on their high-temperature
sides; see figure~\ref{fig:alphaa}~(c). This observation
indicates that in high magnetic fields two separate transitions
have to be distinguished. A natural interpretation is that with
increasing temperature a first-order transition from the LTC to
the LTI phase takes place and this transition is followed by a
second-order transition from the LTI to the HTI phase at a
slightly higher temperature ($\simeq 0.5-1$~K). In this respect,
there is no qualitative difference between the sequence of
transitions in intermediate ($\lesssim 10$~T) and in higher
fields ($\gtrsim 11$~T). This conclusion is in contrast to
previous publications~\cite{kimura05a,kimura03b} which proposed
direct transitions from the HTI to the LTC phase in TbMnO$_3$ for
$H \gtrsim 11$~T.

The N\'{e}el temperature $T_{\rm N}$ shows only a weak increase of
about 2~K in the field range up to 14~T, i.e., the HTI phase is
slightly stabilized with increasing $H||a$. The field dependence
of $T_{\rm kink}$ is closely linked to the critical field
$H^{a}_{\rm FE,C}\simeq 9.5$~T. For $H<H^{a}_{\rm FE,C}$, $T_{\rm
kink}\simeq 33.5$~K is nearly field independent and jumps to
$T_{\rm kink}\simeq 28.5$~K at $H^{a}_{\rm FE,C}$.\footnote{The
kink-like anomaly is only present in measurements of
$\alpha_i(T)$ ($i=a$, $b$, $c$) with increasing temperature. If
$\alpha_i$ is measured with decreasing temperature, no kink
appears.}

\subsection{Measurements in magnetic fields $H||b$}\label{Mn_b}

\begin{figure}
    \begin{center}
        \includegraphics[width=0.90\textwidth]{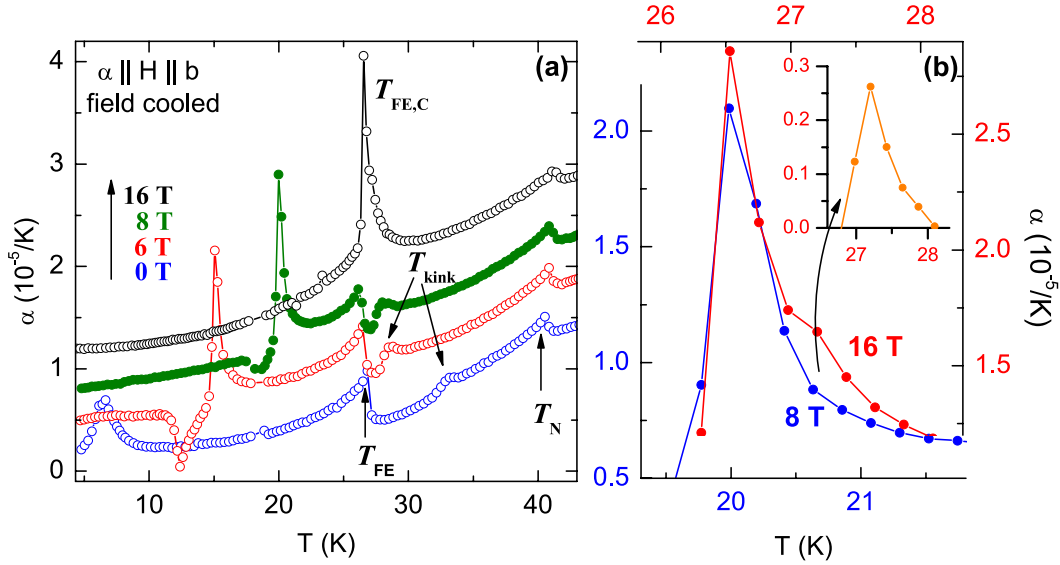}
    \end{center}
    \caption{Panel (a): FC thermal expansion $\alpha_b(T)$ for $H||b$
    measured with increasing temperature. For clarity the curves for
    different fields are shifted by $4\cdot 10^{-6}$/K with respect
    to each other. Panel~(b) compares the anomalies of $\alpha_b(T)$
    at $T_{\rm FE,C}$ measured in $H=8$~T (left and bottom scales) and in
    $H=16$~T (right and top scales). The additional shoulder
    on the high-temperature side of the anomaly in 16~T is
    obvious. As is shown in the inset, the magnitude of this
    shoulder is comparable to the anomalies due to the
    HTI-to-LTI transitions at $T_{\rm FE}$ observed in lower
    fields.}
    \label{fig:alpha_b}
\end{figure}

The thermal expansion coefficient $\alpha_b(T)$ is presented in
figure~\ref{fig:alpha_b}~(a) for some representative magnetic
fields $H||b$. Similar to the measurements in $H||a$, the N\'{e}el
temperature weakly increases with field and the anomaly at $T_{\rm
kink}$ shows a jump-like decrease at the critical field $H_{\rm
FE,C}^{b}\simeq 5$~T. In a magnetic field of 16~T, a pronounced
anomaly at $T_{\rm FE,C}=26.5$~K signals the first-order
LTI-to-LTC transition. Again this anomaly has an additional
shoulder on the high-temperature side, which arises from the
second-order LTI-to-HTI transition at $T_{\rm FE}\simeq 28$~K. For
$H=8$~T these two transitions are well separated from each other,
$T_{\rm FE}\simeq 26$~K and $T_{\rm FE,C}\simeq 20$~K, and their
different order is reflected in different shapes of the respective
anomalies. In figure~\ref{fig:alpha_b}~(b) we compare the
anomalies of $\alpha_b(T)$ at $T_{\rm FE,C}$ for $H=8$~T and
$H=16$~T by shifting the 8~T curve on top of the 16~T curve. No
further scaling is applied. The additional shoulder of the 16~T
curve is clearly seen and a subtraction of the shifted 8~T curve
even allows for a quantitative estimate of its magnitude. As
shown in the inset of figure~\ref{fig:alpha_b}~(b), the additional
shoulder in 16~T is of comparable magnitude as the anomalies due
to the HTI-to-LTI transitions at $T_{\rm FE}$ observed in lower
fields. Thus, our data for $H||b$ allow for the same conclusion
as drawn above from our data for $H||a$. For both field
directions we do not observe direct HTI-to-LTC transitions.

\begin{figure}
    \begin{center}
        \includegraphics[width=0.80\textwidth]{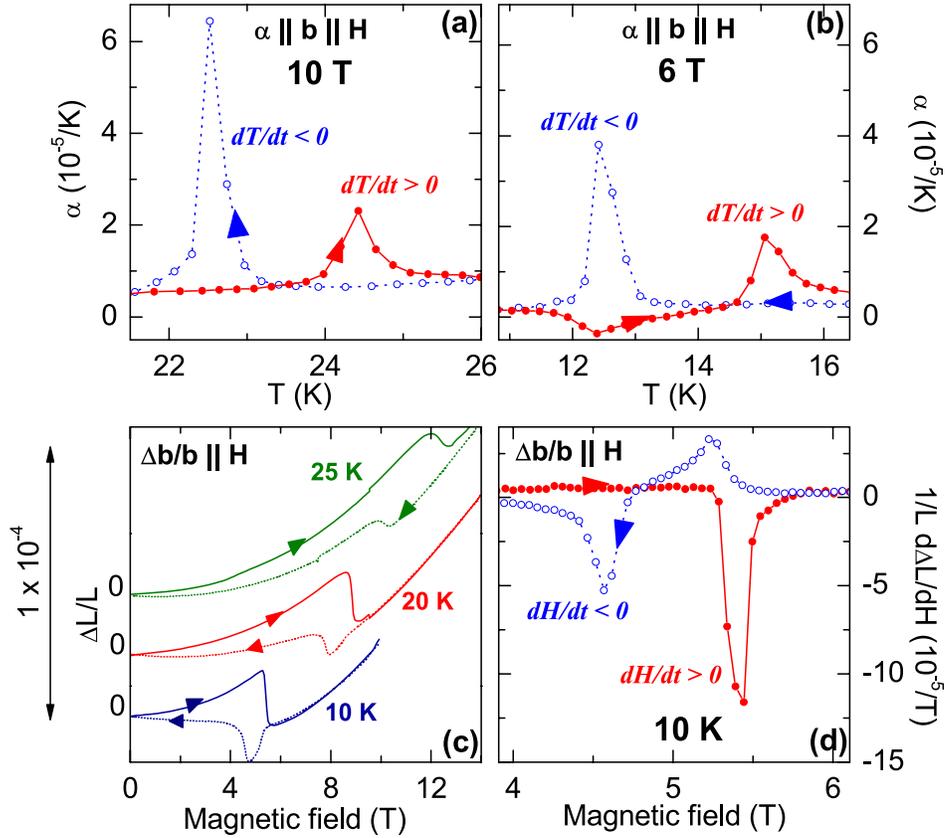}
    \end{center}
    \caption{Hysteresis of the transition between the LTI and the
    LTC phases for $H||b$. Panel (a) and (b) show $\alpha_b(T)$ measured with
    increasing ($\bullet$--) and decreasing ($\circ\cdots$) $T$ in a
    field of 10~T and of 6~T, respectively. In 6~T, the anomalies are
    of completely different shapes for different signs of $dT/dt$, while
    only the widths of the anomalies change in $H=10$~T. Panel (c)
    shows magnetostriction data recorded with increasing (---) and
    decreasing ($\cdots$) magnetic field at different $T$. For
    different signs of the field sweep the anomalies
    only weakly change at 25~K, while they completely change their
    shapes for $T\leq 20$~K. Panel (d) displays the field
    derivatives of the magnetostriction data at $T=10$~K for increasing
    ($\bullet$--) and decreasing ($\circ\cdots$) field. The
    LTI-to-LTC transition causes a single peak, while a double peak is
    observed at the LTC-to-LTI transition, in complete analogy to
    the anomalies of $\alpha_b(T)$ shown in panel~(b).}
    \label{fig:hysteresisb}
\end{figure}

Figure~\ref{fig:hysteresisb}~(a) displays the anomalies of
$\alpha_b(T)$ at $T_{\rm FE,C}$ measured with increasing and
decreasing temperature in $H=10$~T. The shape of the anomalies is
typical for a first-order transition and there is a clear
hysteresis. The different peak heights are not too surprising
because this simply means that not only the value of $T_{\rm
FE,C}$ but also the width of the transition depends on the sign
of the temperature drift. A much more surprising feature is found
for lower fields, as is shown for $H=6$~T in panel~(b). Here, we
observe a double-peak structure when the phase boundary from the
LTC to the LTI phase is crossed with increasing temperature.
However, for the opposite direction, i.e.\ $dT/dt<0$, a single
peak signals the LTI-to-LTC transition. The same anomalous
behavior is observed in our magnetostriction measurements. As
shown in figure~\ref{fig:hysteresisb}~(c), a jump-like contraction
of the $b$ axis at $H\simeq 5.4$~T signals the LTI-to-LTC
transition as a function of increasing $H$ at $T=10$~K. This is
typical for a first-order transition and the corresponding field
derivative of $\Delta b(H)/b$ shows a peak of negative sign; see
figure~\ref{fig:hysteresisb}~(d). On decreasing the field again,
we do not, however, observe the expected jump-like expansion at
the LTC-to-LTI transition. Instead there is even a further {\em
decrease} of $\Delta b(H)/b$ in a restricted field range, which
causes a double-peak in the corresponding field derivative. These
highly anomalous double-peak structures only appear at the
LTC-to-LTI transitions at temperatures below about 20~K and for
$H||b$. In other compounds, double peaks of $\alpha(T)$ have been
observed on heating through glass-like transitions and it has been
found that these peaks depend on the rate of the previous cooling
process~\cite{mueller02a,gugenberger92a}. In order to test
whether such memory effects also exist in TbMnO$_3$, we varied
the rate of the increasing field between 20~mT/min and 2~T/min,
but we did not observe any influence in the subsequent
measurements with decreasing field. At present, the origin of the
anomalous double-peak structures remains unclear.

\subsection{Measurements in magnetic fields $H||c$}

\begin{figure}
    \centering
        \includegraphics[width=0.90\textwidth]{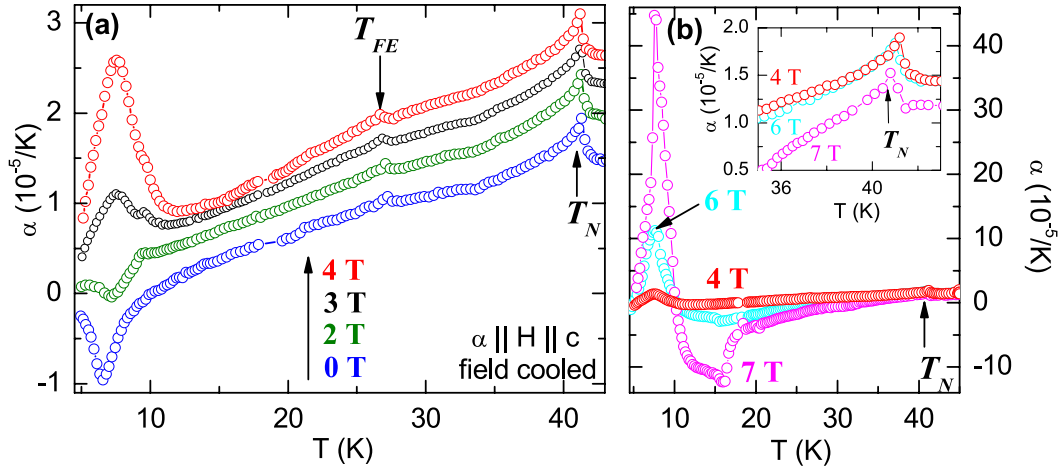}
    \caption{Panel (a): FC thermal expansion $\alpha_c(T)$ for $H||c$
    measured with increasing temperature. For clarity the curves for
    different fields are shifted by $4\cdot 10^{-6}$/K with respect
    to each other. Panel (b) compares the $\alpha_c(T)$ curves (not
    shifted) in the field range between 4 and 7~T and the inset
    shows an expanded view of the anomalies at $T_{\rm N}$.}
    \label{fig:alpha_c}
\end{figure}

According to Ref.~\cite{kimura05a}, no additional anomalies
arising from the Mn sublattice are expected in magnetic fields
along the $c$ axis up to $H\simeq 7$~T. Larger fields induce a
canted AFM ordering of the Mn spins and the ferroelectric
polarization is suppressed. Figure~\ref{fig:alpha_c}~(a) shows
the thermal expansion coefficient $\alpha_c(T)$ in fields up to
4~T and, indeed, these curves do hardly change above about 12~K.
The drastic changes in the lower temperature region arise from the
Tb sublattice and will be discussed in section~\ref{Tbordering}.
Figure~\ref{fig:alpha_c}~(b) presents the $\alpha_c(T)$
measurements in the field range from 4~T and 7~T. In 6~T,
$\alpha_c(T)$ already shows a broad minimum around 15~K, which
changes into a well-pronounced anomaly at $\simeq 16.5$~K in the
$7$~T curve. We attribute this anomaly to the transition from the
LTI to the cAFM phase. Unfortunately, the investigation of this
phase boundary by high-resolution dilatometry was not possible at
higher fields for $H||c$ because of strong torque effects.
Usually, the sample is clamped in the dilatometer by only a very
small pressure, which was, however, not sufficient to prevent a
rotation of the sample when the phase boundary of the cAFM phase
was crossed. Thus, a fitting was constructed to safely fix the
sample orientation, but the torque effects were sufficiently
strong to crack the sample in this setup.

\section{$(H,T)$-phase diagram and pressure dependencies}

\begin{figure}
    \begin{center}
        \includegraphics[width=0.85\textwidth]{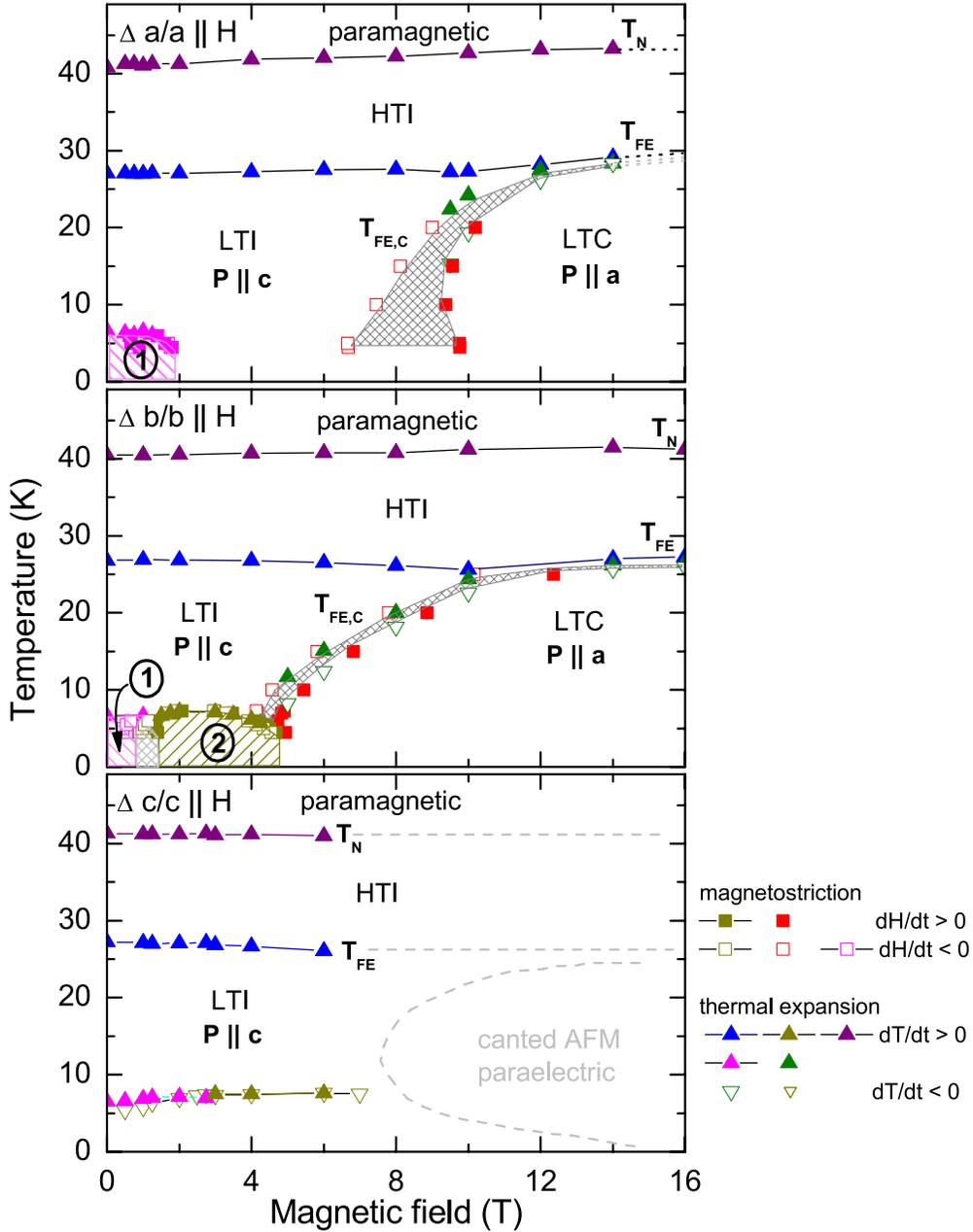}
    \end{center}
    \caption{Phase diagram of TbMnO$_3$ based on the measurements
    of $\alpha_i(T)$ (triangles) and $\Delta L_i(H)/L_i$ (squares) measured
    along $i=a$, $b$, $c$ with $H||i$. Filled and open symbols are obtained with
    increasing and decreasing temperature and field, respectively. The dashed
    lines depict the high-field phase boundaries obtained in
    Ref.~\cite{kimura05a}, because we could measure our sample only up to 7~T
    for $H||c$ (see text).}
    \label{fig:Phasendiagramm}
\end{figure}

From our thermal expansion and magnetostriction measurements we
obtain the phase diagram shown in figure~\ref{fig:Phasendiagramm}.
The phase boundary between the LTI and the LTC phase, measured
as a function of increasing and decreasing field $H||a$ and temperature, has been
determined for the first time. We find a very strong hysteretic
behavior, especially at low temperatures. Concerning the phase
boundaries between the paramagnetic and the HTI phases our data
well agree to previous investigations~\cite{kimura03b,kimura05a,aliouane05a} for
all three field directions. In the field range below about
10~Tesla, this is also the case for the boundaries between the
HTI and LTI phases. In contrast to the earlier publications,
however, our data clearly show that there are no direct
transitions from the HTI to the LTC phase in higher fields,
neither for $H||a$ nor for $H||b$. Instead of a direct transition
from the HTI to the LTC phase as a function of decreasing
temperature, the system seems to pass always through the LTI
phase, before the LTC phase is entered. This observation is of
importance for microscopic models describing the symmetry changes
at the various phase transitions.

Besides the bare position of the different phase boundaries, the
thermal expansion data also yield information about their uniaxial
pressure dependencies (see eq.~(\ref{Ehr})). As discussed in
detail in Ref.~\cite{baier06b}, the opposite signs of the
anomalies of $\alpha_a$ and $\alpha_b$ at both, $T_{\rm N}$ and
$T_{\rm FE}$, mean that both transitions depend on the degree of
distortion of the GdFeO$_3$-type structure, which can be
parameterized by the magnitude of the orthorhombic splitting
$\varepsilon=(b-a)/(b+a)$. The main idea is that the
ferromagnetic nearest-neighbor exchange $J_{\rm NN}^{\rm FM}$ in
the $ab$ planes is weakened with increasing $\varepsilon $, as it
is the case in the $R$MnO$_3$ series for a decreasing ionic
radius from $R={\rm La}\dots {\rm Dy}$. In addition, the
anisotropy of the antiferromagnetic next-nearest-neighbor
coupling $J_{\rm NNN}^{\rm AFM}$ increases, since $J_{\rm
NNN}^{\rm AFM}$ increases along $b$ and decreases along $a$. As a
consequence, the zero field magnetic ground state changes from an
A-type AFM for $R={\rm La}\dots {\rm Gd}$ to an E-type AFM for
the HoMnO$_3$ in perovskite structure.\footnote{$R$MnO$_3$ with
rare earth ions smaller than Dy usually crystallize in a
hexagonal structure\cite{goodenough06a,lottermoser01a}} Depending
on temperature, magnetic field and the radius of $R$, the
competition of $J_{\rm NN}^{\rm FM}$ and $J_{\rm NNN}^{\rm AFM}$
can also cause incommensurate AFM structures for $R={\rm Eu}\dots
{\rm Ho}$~\cite{goto04a,arima05a}. Since uniaxial pressure along
$a$ ($b$) will increase (decrease) $\varepsilon $, the
corresponding uniaxial pressure dependencies of $T_{\rm N}$ of
TbMnO$_3$ can be straightforwardly traced back to a decreasing
(increasing) $J_{\rm NN}^{\rm FM}$ due to the changes of the
orthorhombic splitting. This result is identical to our
conclusions concerning the uniaxial pressure dependencies of
$T_{\rm N}$ of GdMnO$_3$ for $p_a$ and $p_b$. For uniaxial
pressure $p_c$, we find an increase of $T_{\rm N}$ in both
compounds. As discussed in Ref.~\cite{baier06b}, the
GdFeO$_3$-type distortion is characterized not only by the
magnitude of $\varepsilon $, but also by a decreasing lattice
parameter $c$. Thus, one might expect a negative $\partial T_{\rm
N} / \partial p_c$ arising from a partial increase of the
GdFeO$_3$-type distortion lowering $J_{\rm NN}^{\rm FM}$. However,
$T_{\rm N}$ does not only depend on the couplings $J_{\rm
NN}^{\rm FM}$ and $J_{\rm NNN}^{\rm AFM}$ acting within the $ab$
planes. A three-dimensional ordering requires a finite coupling
$J_{c}^{\rm AFM}$ along the $c$ direction, and the positive
$\partial T_{\rm N} / \partial p_c$ suggests an increase of
$J_{c}^{\rm AFM}$ under pressure along $c$.

Concerning the HTI-to-LTI phase boundary and its pressure
dependencies, the phenomenology of TbMnO$_3$ is not directly
comparable to our previous results on
GdMnO$_3$~\cite{baier06a,baier06b}. In TbMnO$_3$, the transition
to the LTI phase is accompanied by a finite polarization already
in zero magnetic field, whereas in GdMnO$_3$ a finite field
$H||b$ is necessary to induce ferroelectricity. In addition, the
ferroelectric phase of GdMnO$_3$ is not entered directly from the
the paraelectric high-temperature incommensurate phase (termed
ICAFM in Refs.~\cite{baier06a,baier06b}) but from a canted AFM
phase.\footnote{Note that the magnetic structures of GdMnO$_3$
have not yet been determined unambiguously. The proposed phases
are based on the observed weak
ferromagnetism~\cite{kimura03b,hemberger04b} and on X-ray
diffraction studies~\cite{arima05a}. Magnetic neutron diffraction
is still missing.} Concerning the uniaxial pressure dependencies
for $p_a$ and $p_b$, we observed a clear anti-correlation in
GdMnO$_3$ between $T_{\rm N}$ and $T_{\rm c}$ (the latter signals
the HTI-to-cAFM boundary) on the one hand side and $T_{\rm FE}$
on the other. This anti-correlation suggests a competition
between colinear spin structures, either incommensurate or A-type
AFM, and non-colinear spin structures, which allow for additional
ferroelectricity~\cite{mostovoy06a,katsura05a}. In contrast to
GdMnO$_3$, the uniaxial pressure dependencies of $T_{\rm N}$ and
$T_{\rm FE}$ of TbMnO$_3$ have the same signs. An increase of the
orthorhombic splitting by pressure would decrease both, $T_{\rm
N}$ and $T_{\rm FE}$. This is in agreement with the observed lower
values of both, $T_{\rm N}$ and $T_{\rm FE}$ of DyMnO$_3$ where
the orthorhombic splitting is larger than in
TbMnO$_3$~\cite{goto04a,arima05a}. Our finding suggests that the
orthorhombic splitting of TbMnO$_3$ is already larger than the
optimum value of $\varepsilon $ needed to establish a
ferroelectric ordering with a maximum value of $T_{\rm FE}$.
Summarizing the above discussion, we conclude that the
competition between a multiferroic and an A-type AFM ground state
dominates in GdMnO$_3$, while in TbMnO$_3$ the dominant
competition is between the multiferroic and an E-type AFM ground
state.

Apart from the phase boundaries related to the Mn subsystem,
figure~\ref{fig:Phasendiagramm} also contains various phase
boundaries in the low-temperature range arising from transitions
of the Tb ions. The corresponding measurements, which have been
used to trace these phase boundaries will be discussed in the
following section. We find evidence that only one ordered phase
of the Tb moments exists for $H||a$, while for $H||b$ it is
possible to distinguish at least two different phases. For $H||c$
the situation is more complex because a clear separation of
different phases is not possible from our data.

\section{Transitions related to the Tb ions}\label{Tbordering}

Even though the Tb ions have a strong magnetic moment which may
interact with the Mn moments, no systematic investigation on the
field dependence of the low-temperature phase transition has been
published so far. According to
Refs.~\cite{kajimoto04a,aliouane05a} an incommensurate AFM
ordering of the Tb moments occurs in zero field below $T_{\rm
N}^{\rm Tb}=7$~K. In Ref.~\cite{kimura03b} Kimura {\it et al.}
present several phase boundaries in the relevant temperature
region for $H||b$ only and without a further classification.
Therefore, we have studied this low-temperature region in more
detail for all three field directions.

\subsection{Measurements in magnetic fields $H||a$}

\begin{figure}
    \begin{center}
        \includegraphics[width=0.90\textwidth]{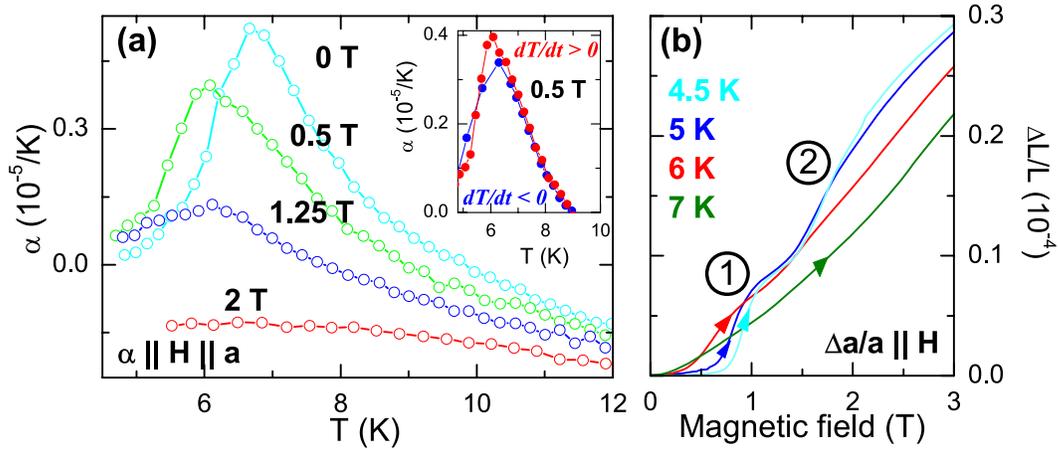}
    \end{center}
    \caption{Panel (a): FC thermal expansion $\alpha_a(T)$ for
    $H||a$ showing the suppression of the Tb ordering above about
    2~T.
    Panel (b): Magnetostriction measurements $\Delta a(H)/a$ recorded
    with increasing field below $T_{\rm N}^{\rm Tb}$.}
    \label{fig:Tb_a}
\end{figure}

As shown in figure~\ref{fig:Tb_a} (a), a positive anomaly of
$\alpha_a(T)$ at $T_{\rm N}^{\rm Tb}\simeq 7$~K signals the Tb
ordering in zero field. This anomaly shifts to lower temperature
as small magnetic fields are applied and it is completely
suppressed for $H \geq 2$~T. The suppression of the
antiferromagnetic order is in agreement with the ferromagnetic
alignment of the Tb moments for $H\geq 2$~T reported in
Ref.~\cite{aliouane05a}. Our data suggest that the Tb ordering is
of second order since no hysteretic behavior occurs at the phase
boundary (see the inset of figure\ref{fig:Tb_a}~(a)). This phase
transition can also be detected by magnetostriction measurements.
The $\Delta a(H)/a$ curves in figure~\ref{fig:Tb_a}~(b) were
obtained with increasing $H$ at constant $T$. Considering $\Delta
a(H)/a$ at 4.5~K, a first step-like expansion occurs at $H=0.9$~T
and a second one at $H=1.7$~T. The second one can be attributed
to the ferromagnetic alignment of the Tb
moments~\cite{aliouane05a}, but the origin of the first anomaly
remains unclear thus far. Again, both anomalies show no
hysteretic behavior as a function of field (not shown). With
increasing temperature these anomalies broaden and vanish for
$T>T_{\rm N}^{\rm Tb}\simeq 7$~K.

\subsection{Measurements in magnetic fields $H||b$}

\begin{figure}
    \begin{center}
        \includegraphics[width=0.90\textwidth]{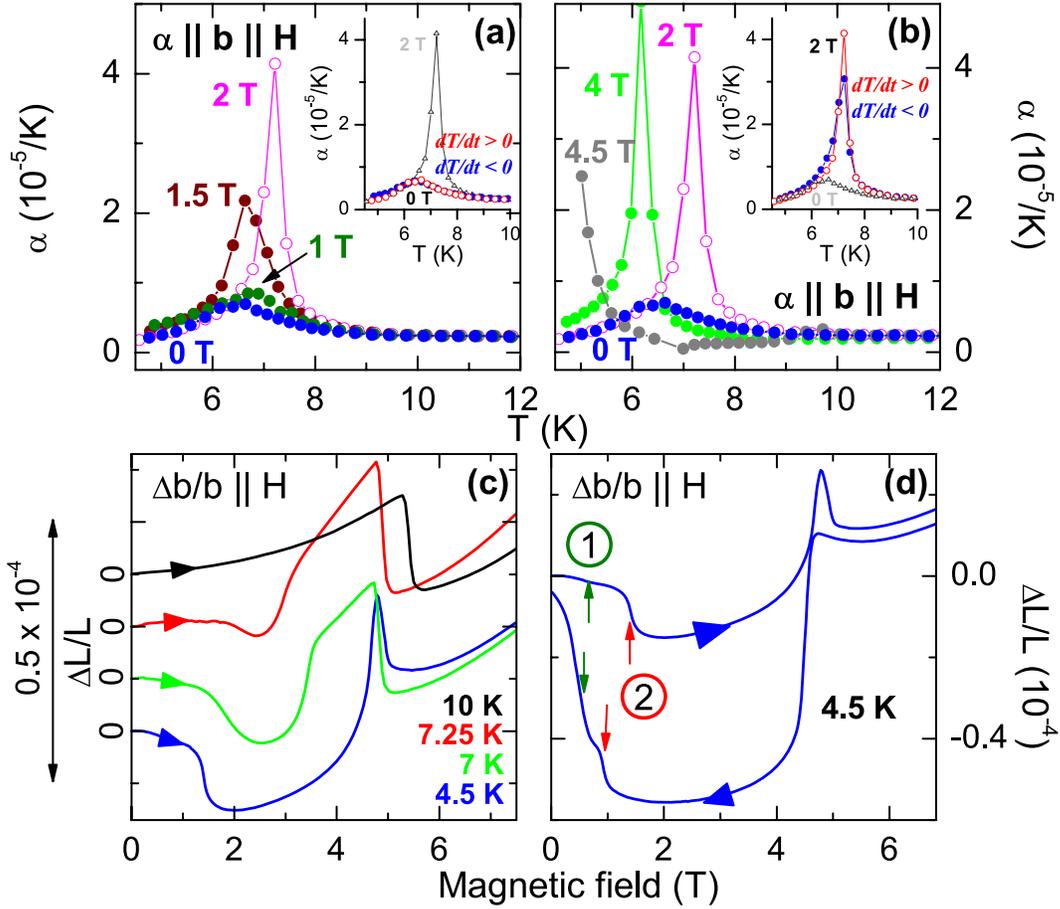}
    \end{center}
    \caption{Panel (a) and (b): FC thermal expansion $\alpha_b(T)$ for $H||b$
    showing two different types of anomalies at $T_{\rm N}^{\rm Tb}$. Below about 1~T
    the anomaly is rather broad and hardly changes with field. For larger
    fields the anomaly becomes very sharp and its position changes
    non-monotonically with field. Panel~(c): Magnetostriction $\Delta b(H)/b$
    recorded with increasing field. For clarity, the curves are offset with
    respect to each other. Panel~(d): Magnetostriction $\Delta b(H)/b$
    recorded with increasing and decreasing field at $T=4.5$~K (see text).}
    \label{fig:Tb_b}
\end{figure}

For magnetic fields $H||b$ up to $H\simeq 1$~T, $\alpha_b(T)$
shows similar anomalies at $T_{\rm N}^{\rm Tb}$ as those observed
in $\alpha_a(T)$; see figures~\ref{fig:Tb_b}~(a)
and~\ref{fig:Tb_a}~(a). In contrast to $\alpha_a(T)$, however,
the antiferromagnetic Tb ordering is not suppressed as $H$ is
increased. Instead, another ordered phase is induced for $H >
1$~T. The anomalies in higher fields are sharper and
significantly larger than those in $H \leq 1$~T. For fields
$H||b$ up to 2~T this second Tb ordering is stabilized, but with
further increasing field the transition temperature decreases
again and at 4.5~T the transition already occurs below the lower
limit of the investigated temperature region, see
figure~\ref{fig:Tb_b}~(b). The insets of panel~(a) and (b) of
figure \ref{fig:Tb_b} compare measurements of $\alpha_b(T)$
obtained with increasing and decreasing $T$. For both types of
transitions no hysteresis is observed. In Ref.~\cite{aliouane05a}
it has been reported that the wave vector of the antiferroamgnetic
Tb oredering changes from an incommensurate value below 1~T to a
commensurate one for higher fields $H||b$. Thus we attribute the
broad low-field anomalies in $\alpha_b(T)$ to transitions to the
incommensurate phase, while the sharper anomalies above 1~T signal
transitions to the commensurate phase.

Panel~(c) of figure~\ref{fig:Tb_b} presents the relative length
change $\Delta b(H)/b$ as a function of increasing $H$. These
curves confirm that two different phase transitions have to be
distinguished in $\alpha_b(T)$ at $T_{\rm N}^{\rm Tb}$. As a
function of increasing $H$, $\Delta b(H)/b$ shows a pronounced
step-like contraction at 1.4~T, in agreement with the observed
transition from an incommensurate to a commensurate wave vector
of the Tb ordering~\cite{aliouane05a}. At a somewhat larger field
we find an anomalous expansion of the $b$ axis suggesting that
the commensurate phase is left again. With further increasing
field another step-like contraction takes place, which is due to
the LTI-to-LTC transition of the Mn moments discussed in
section~\ref{Mn_b}. The latter is present up to about 25~K (see
figure~\ref{fig:hysteresisb}), while the transitions at lower
fields can be observed only at temperatures up to $T\simeq
7.25$~K. Thus, the transition temperature of the commensurate Tb
phase is slightly higher than $T_{\rm N}^{\rm Tb}$ in zero field.

Figure~\ref{fig:Tb_b}~(d) shows an expanded view on $\Delta
b(H)/b$ at 4.5~K as a function of increasing and decreasing field.
As in $\Delta a(H)/a$, there is an additional anomaly (of unknown
origin) marked by \ding{192}, before the
incommensurate-to-commensurate transition occurs at the anomaly
marked by \ding{193}. The position of the first anomaly is not
hysteretic with respect to the direction of the field sweep,
whereas the position of the second one as well as the anomalies
around 5~Tesla show some hysteresis. Our phase boundaries
qualitatively agree with the phase diagram presented by Kimura
{\it et al.} for $H||b$ in Ref.~\cite{kimura03b}.

\subsection{Measurements in magnetic fields $H||c$}

\begin{figure}
    \begin{center}
        \includegraphics[width=0.70\textwidth]{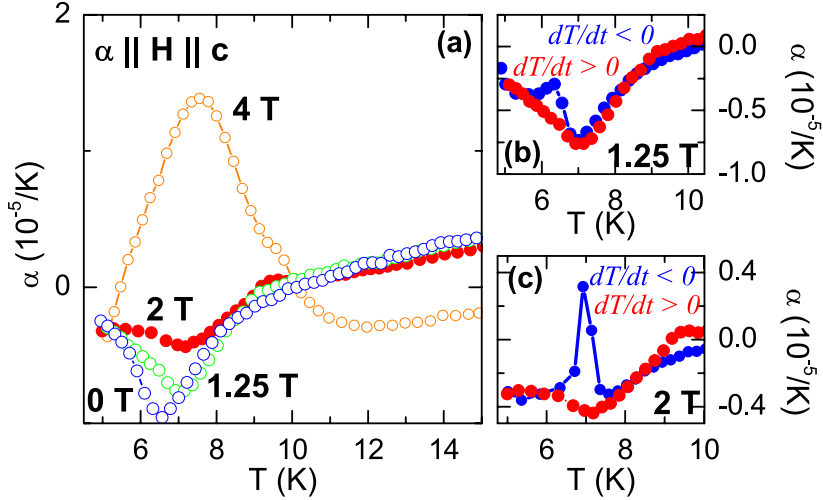}
    \end{center}
    \caption{Panel (a): FC thermal expansion $\alpha_c(T)$ for $H||c$
     around $T_{\rm N}^{\rm Tb}$. Panel (b) and (c): Thermal expansion recorded
     with increasing ($\bullet$) and decreasing (-$\bullet$-) temperature.}
    \label{fig:Tb_c}
\end{figure}

For a magnetic field $H||c$, we find a complex field dependence
of the anomalies of $\alpha_c(T)$, which signal the various
rearrangements of the Tb moments below $T_{\rm N}^{\rm Tb}$. In
zero field, an anomaly of negative sign indicates the ordering of
the Tb sublattice. This anomaly broadens as higher fields are
applied and can be observed up to $\simeq 2$~T; see figure
\ref{fig:Tb_c}~(a). For higher fields, a broad anomaly of
positive sign shows up. This anomaly continuously increases in
magnitude for further increasing $H$ and has been detected up to
$H=7$~T; see also figure~\ref{fig:alpha_c}~(b). Presumably, the
two anomalies of opposite signs belong to two types of different
Tb phases similar to those observed for $H||b$. The need to
distinguish two anomalies also for $H||c$ becomes more obvious in
panel~(b) and~(c) of figure~\ref{fig:Tb_c}. There, we present the
anomalies of $\alpha_c(T)$ recorded with increasing and
decreasing temperature in $H=1.25$~T and $H=2$~T. For both field
strengths, $\alpha_c(T)$ shows only one broad anomaly of negative
sign as a function of increasing $T$. For decreasing $T$,
however, this anomaly is overlapped by a second one of opposite
sign. This complex hysteretic behavior prevents an exact
determination of clearly-defined phase boundaries. Nevertheless,
our data suggest that the Tb ordering is not suppressed up to 7~T.

\section{Summary}

We have determined the magnetic-field temperature phase diagram of
TbMnO$_3$ by high-resolution thermal expansion and
magnetostriction measurements. The measurements have been
performed in longitudinal magnetic fields applied along all three
crystallographic axes. The fact that we find rather pronounced
anomalies at the various phase transitions proves that all these
transitions strongly couple to lattice degrees of freedom and
allows for a detailed investigation of the phase boundaries. Our
data reveal various new features in the phase diagram of
TbMnO$_3$: Firstly, the phase boundary between the LTI and the
LTC phase for $H||a$ has been determined for the first time. This
first-order phase transition shows a broad hysteresis, which is
strongly enhanced at lower temperatures. Secondly, we find clear
evidence that even in high magnetic fields $H||a$ no direct
HTI-to-LTC transitions take place, contrary to previous
reports~\cite{kimura05a,kimura03b}. This is also the case for
high fields $H||b$. Our data suggest that for both field
directions the Mn subsystem always transform from the HTI first
to the LTI phase before the LTC phase is finally established at
low temperature. Thirdly, we observe a strongly anomalous
behavior at the LTC-to-LTI transition for $H||b$. Similar to
glass transitions, double-peak structures show up in
$\alpha_b(T)$ and $\frac{\partial \Delta L/L}{\partial H}$ when
the LTC phase is left as a function of temperature and magnetic
field, respectively. Nevertheless, we could not observe any
dependence of these double-peak structures on the sweep rate of
the temperature or magnetic-field changes during the prior
LTI-to-LTC transitions, which would be a typical indication for a
glass-like transition.

Besides the positions of the various phase boundaries, our data
also yield information about their uniaxial pressure dependencies.
The uniaxial pressure dependencies of the N\'{e}el temperature $T_N$
of TbMnO$_3$ have the same signs as those of GdMnO$_3$. This
confirms our previous conclusions~\cite{baier06b}, that the
increase (decrease) of $T_N$ for uniaxial pressure applied along
the $a$ ($b$) axes arises from a decrease (increase) of the
orthorhombic splitting $\varepsilon$, which causes an increase
(decrease) of $J_{\rm NN}^{\rm FM}$. In order to explain the
uniaxial pressure dependence for pressure along the $c$ axis, the
finite $J_{c}^{\rm AFM}$ has to be taken into account. The
analysis of the pressure dependencies of $T_{\rm FE}$ suggests
that the optimum value of $\varepsilon$ needed to establish a
ferroelectric order with a maximum $T_{\rm FE}$ is located between
$\varepsilon$ of GdMnO$_3$ and $\varepsilon$ of TbMnO$_3$.

Concerning the ordering of the Tb moments, our data confirm the
suppression of the incommensurate ordering below $T_{\rm N}^{\rm
Tb}=7$~K for magnetic fields $H \gtrsim 2$~T applied along the
$a$ axis. For $H||b$, we also observe that a clear change in the
ordering of the Tb moments is induced for $H \gtrsim 1$~T, in
agreement with the incommensurate-to-commensurate transition
found by neutron scattering~\cite{aliouane05a}. Clear anomalies,
due to the ordering of the Tb sublattice are also present for
$H||c$ and our data indicate a rearrangement of the Tb moments
around 3~T. The ordering is not suppressed in fields up to 7~T,
but a clear attribution to different phases is prevented by the
very complex, hysteretic field and temperature dependencies.

\section*{Acknowledgements}

We acknowledge fruitful discussions with J.~Baier, K.~Berggold,
J.~Hemberger, and D.~Khomskii. This work was supported by the
Deutsche Forschungsgemeinschaft via Sonderforschungsbereich 608.

\section*{References}
\bibliographystyle{unsrt}

\end{document}